\documentclass[aps,prb,reprint,showpacs,superscriptaddress,groupedaddress]{revtex4-1}
\usepackage{graphicx}
\usepackage{amsmath}
\usepackage{amssymb}
\usepackage{dcolumn}
\usepackage{dsfont}
\usepackage{latexsym}
\usepackage{rotating}
\usepackage{hyperref}
\usepackage{amsmath,amssymb,amsfonts}
\usepackage{bm}
\usepackage{color}
\usepackage{graphicx}
\usepackage{amsmath}
\usepackage{amssymb}
\usepackage{dcolumn}
\usepackage{dsfont}
\usepackage{latexsym}
\usepackage{rotating}
\usepackage{bbm}
\usepackage[usenames,dvipsnames]{xcolor}  
\usepackage{float}
\usepackage{epsfig} 
\usepackage{psfrag}
\usepackage{natbib}
\usepackage{bm}
\usepackage{eucal}
\usepackage{mathrsfs}
\usepackage{braket}
\usepackage{enumerate}
\usepackage{longtable}
\usepackage{subfigure}
\usepackage{bm}
\usepackage{hyperref}
\usepackage{amsfonts}
\usepackage{dcolumn}
\usepackage{bm}
\usepackage{subfigure}

\newcommand{\be}{\begin{equation}}
\newcommand{\ee}{\end{equation}}
\newcommand{\bn}{\begin{eqnarray}}
\newcommand{\en}{\end{eqnarray}}

\usepackage{color} 

\usepackage{hyperref}
\hypersetup{
colorlinks=true,final=true,
        linkcolor=red,
        citecolor=blue,
        filecolor=blue,
        urlcolor=blue,
}
\begin{document}
\noindent

\title{Emergent Strange Nodal Metallicity from Orbital-Selective Mott Physics}
\author{Swagata Acharya$^{1,2}$}\email{swagata.acharya@kcl.ac.uk}
\author{Mukul S. Laad$^{3}$}\email{mslaad@imsc.res.in}
\author{Nagamalleswararao Dasari$^{4}$}\email{nagamalleswara-rao.dasari@mpsd.mpg.de}
\author{N. S. Vidhyadhiraja$^{5}$}\email{raja@jncasr.ac.in}
\author{Mark Jarrell$^{6,7}$}\email{jarrellphysics@gmail.com}
\author{A. Taraphder$^{2,8}$}\email{arghya@phy.iitkgp.ernet.in}
\affiliation{$^{1}$King's College London, Theory and Simulation of Condensed Matter, The Strand, London WC2R 2LS, UK}
\affiliation{$^{2}$Department of Physics, Indian Institute of Technology,
Kharagpur, Kharagpur 721302, India.}
\affiliation{$^{3}$Institute of Mathematical Sciences, Taramani, Chennai 600113, India
}
\affiliation{$^{4}$ Max Planck Institute for the Structure and Dynamics of Matter,  22761 Hamburg,Germany}
\affiliation{$^{5}$Theoretical Sciences Unit, Jawaharlal Nehru Centre For Advanced Scientific Research, Jakkur, Bangalore 560064, India.}
\affiliation{$^{6}$Department of Physics $\&$ Astronomy, Louisiana State University, Baton Rouge, LA 70803, USA}
\affiliation{$^{7}$Center for Computation $\&$  Technology, Louisiana State University, Baton Rouge, Louisiana 70803, USA.}
\affiliation{$^{8}$Centre for Theoretical Studies, Indian Institute of
Technology Kharagpur, Kharagpur 721302, India.}

\begin{abstract}
	While a specific kind of strange metal is increasingly found to be the ``normal'' states in a wide variety of unconventional superconductors, its microscopic origin is presently a hotly debated enigma.  
	Using dynamical mean-field theory (DMFT) based on hybridization expansion of continuous-time quantum Monte-Carlo (CTQMC) solver for a extended  two-band Hubbard model (2BHM), 
	we investigate the conditions underlying the emergence of such a metal.  
	Specifically, we tie strange metallicity to an orbital-selective Mottness in 2BHM or 
	momentum-selective Mott phase (OSMP) in 2D Hubbard models inspired by a cluster-to-orbital 
	mapping. We find $(i)$ disparate spin and charge responses, $(ii)$ fractional power-law behavior and $\omega/T$-scaling in the charge and spin fluctuation responses, and $(iii)$ very good 
   accord with optical conductivity and nuclear magnetic relaxation rates in the slightly underdoped normal states of cuprates and Fe-arsenides. We analyze the local problem using bosonization to show that such 
anomalous responses arise from a lattice orthogonality catastrophe specifically in the OSMP.  Our work establishes the intimate link between strange metallicity and selective Mottness in quantum matter.
\end{abstract}


\maketitle

Understanding the strange metal (SM) phase in quantum matter is by now a fundamental problem in electron 
theory of metals. Experimentally, strange metals are increasingly seen as normal states of
unconventional superconductors (USC), such as high-$T_{c}$ in hole-doped cuprates, Iron arsenides ($FeAs$), but also in 
three-dimensional correlated systems close to (partial or complete) Mott localization. In hole-doped cuprates USC maximizes far from an 
anti-ferromagnetic quantum critical point(AF-QCP), but 
close to an optimal doping where a topological Fermi surface reconstruction (FSR), 
connected to a momentum-selective Mott criticality in cluster-DMFT studies~\cite{civelli}, occurs.  In contrast, both these features 
in $FeAs$ and some $f$-electron compounds are found close to an AF-QCP. Very interestingly however, a large to small Fermi surface
reconstruction (FSR) is found in all cases. Thus, this last observation links onset of strange metallicity to a momentum- (in cuprates) 
or orbital-selective Mottness, where such an FSR obtains on quite general grounds. Thus, proximity to AF-QCP may not be a (uniquely)
necessary requirement for generating the soft electronic glue that results in high $T_{c}$ superconductivity (HTSC). This spawns a set of major 
issues: what is the microscopic origin of the anomalous fluctuation spectra in strange metals?  What is its specific link to equally anomalous 
fermiology and transport?  Specifically, the unique signatures of strange metals, generic across a class of systems, are: 
$(i)$ fractional exponents and $\omega/T$-scaling in inelastic neutron scattering (INS)~\cite{aeppli} and 
almost $T$-independent spin relaxation rate in nuclear magnetic resonance (NMR).  
$(ii)$ Quasilinear-in-$T$ resistivity without saturation up to high temperatures and $(iii)$ unusual low-energy optical response, 
$\simeq \omega^{-\eta}$ with $0.7\le \eta < 1.23$~\cite{vdm,akrap}. 
Theoretical rationalizations of these findings contrast strongly, 
ranging from Anderson's hidden-FL~\cite{pwa}, to quantum
phase transitions (QPTs) associated with proximity to various 
($T=0$) ordered phases~\cite{qcp} or, for cuprates, the momentum-selective Mott phases. 

These experimental findings and the above issues motivate this work. We show that the extinction of Landau Fermi Liquid (LFL) quasiparticles 
in an orbital-selective Mott phase (OSMP) can provide a natural understanding of these unique features solely as a consequence of the dualistic
(itinerant-localized) character of carriers. Using dynamical mean-field theory (DMFT), we demonstrate that 
strange metallic behavior naturally {\it emerges} as a consequence of the seminal 
lattice orthogonality catastrophe due to scattering between selectively Mott-localized and metallic states in the OSMP.  We show how a range of spectral and magnetic fluctuation data find rationalization within this novel view, and  explore possible links of our results to Anderson's' {\it hidden Fermi liquid theory}~\cite{pwa} proposal.

\noindent We begin with the 2-band extended Hubbard model (2BHM) (designated by $a,\, b$)~\cite{pepin,laad} 
model, 
\be
H=H_{band} + H_{int} + H_{hyb}
\ee
\noindent where
\vspace{0.3cm}

$H_{band}=\sum_{k,\sigma}\epsilon_{b,k}b_{k,\sigma}^{\dag}b_{k,\sigma} + \sum_{k,\sigma}\epsilon_{a,k}a_{k,\sigma}^{\dag}a_{k,\sigma}$, 
$H_{int}=U_{aa}\sum_{i}n_{ia\uparrow}n_{ia\downarrow} + U_{bb}\sum_{i}n_{ib\uparrow}n_{ib\downarrow} + \sum_{i}U_{ab}n_{ia}n_{ib}$ and 
$H_{hyb}=\sum_{k}V_{ab}(k)(a_{k}^{\dag}b_{k}+h.c)$.  
\vspace{0.3cm}

\noindent Here, $\epsilon_{a,k}=2t_{a}(cosk_{x}+cosk_{y}), \epsilon_{b,k}=2t_{b}(cosk_{x}+cosk_{y}), V_{ab,k}=t_{ab}(cosk_{x}-cosk_{y})$. We 
choose $t_{a}=1.0$~eV, $t_{b}=0.1$~eV and $t_{ab}=0.2$~eV, while $U_{aa,bb}, U_{ab}$ are intra- and inter-orbital Coulomb interactions. 
In high-T$_{c}$ cuprate literature, two-band models very similar to Eq.(1) have a long history. Originally proposed~\cite{varma} in the 
context of marginal-FL theory, they also arise from ab-initio quantum chemical (QC) calculations~\cite{liviu} as effective models for nodal
and anti-nodal states~\cite{imada}, in the context of Mott transitions driven by apical Oxygen displacements~\cite{swagata} and for the hidden order in the pseudogap phase of underdoped cuprates~\cite{giamarchi}. 
Interestingly, a (mathematically) related two-orbital Hubbard model has been derived from a $4$-site  embedded cluster~\cite{werner}. Here, the two orbitals are associated with cluster-adapted bonding and anti-bonding fermionic states arising from the diagonal sites for a one-band Hubbard model on a $4$-site cluster.  This allows a recasting of cluster-DMFT approaches and momentum selective Mott phases in the one-band Hubbard model in terms of orbital-selective 
Mott phases (OSMP) in multi-orbital models.  Multi-orbital Hubbard or Anderson lattice models are natural to $FeAs$~\cite{raghu} or $f$-electron~\cite{hewson} systems.   
\vspace{0.2cm}

\begin{figure}[h!]
\centering
\includegraphics[angle=0,width=0.8\columnwidth]{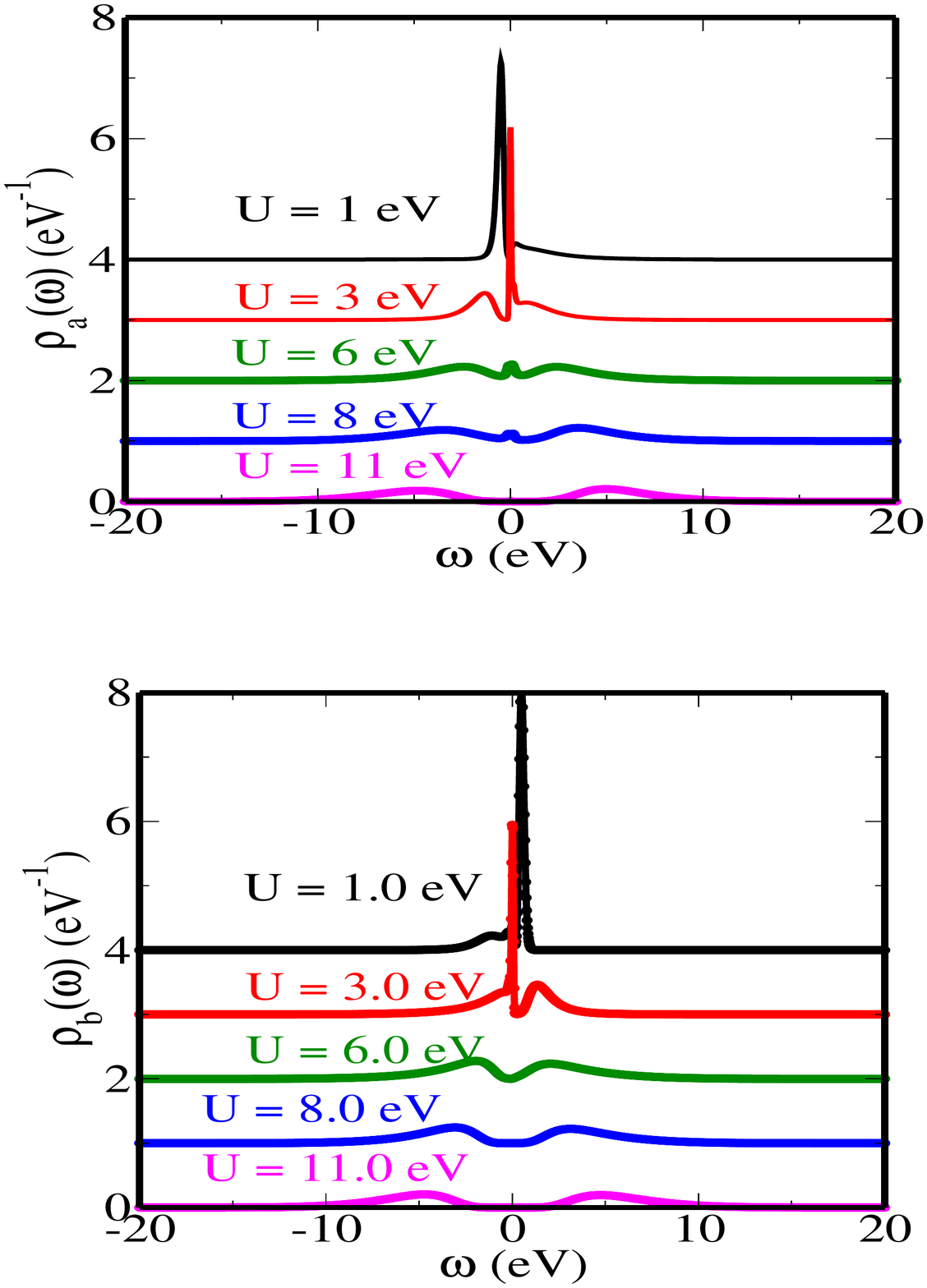}
\caption{The local density of states (DOS) of $a$ and $b$ orbitals are shown at $58$~K and for a range of interaction parameter $U$ varying between $1$~eV and $11$~eV (the DOS at different U are presented with offset of 4, 3, 2, 1 and 0  (eV$^{-1}$) respectively for U=1, 3, 6, 8, 11 eV). The non-interacting DOS has a hybridization gap due to the $D-$~wave form factor of the inter-orbital hybridization. A quasi-particle spectrum appears for $U$ beyond $2.5$~eV. On increasing $U$, first an OSMT and finally, for $U > 11$~eV, a complete Mott gap results.}
\label{dos}
\end{figure}
\vspace{0.2em}

Adopting a model-based approach, we take $U_{aa}=U_{bb}$ and $U_{ab} = 0.3\, U_{aa}$ as interaction parameters.  Clear ${\bf k}$-space 
anisotropy in the Brillouin zone, arising from the ($d$-wave) momentum-dependent hybridization already characterizes the non-interacting 
band structure. Hence, nodal-antinodal separation is reflected in the Fermi surface topology, and the non-interacting local density-of-states 
(LDOS) shows a low-energy hybridization-induced gap. To investigate interaction effects, we solve $H$ (Eq.1) within DMFT using the 
hybridization expansion (CT-hyb) of continuous-time quantum Monte Carlo (CT-QMC) solver~\cite{CT-HYB} in ALPS~\cite{ALPS}.   Now, local one- and two-particle dynamical responses can be 
reliably computed, in contrast to diagrammatic solvers like iterative perturbation theory (IPT), where two-particle susceptibilities need 
reliable knowledge of the fully dynamical but local irreducible vertex. This is presently a demanding task, unless they can be argued to be 
irrelevant, as in large-$N$ theories~\cite{sachdev-ye}.  Using these, we numerically analyze $(i)$ the one- and two-particle spectra for $H$ above,
$(ii)$ its quantum-critical response in the OSMP, and $(iii)$ magnetic fluctuations and optical response in very good accord with candidate strange metals.

\vspace{0.2em}
\begin{figure}[h!]
\centering
\includegraphics[angle=0,width=0.8\columnwidth]{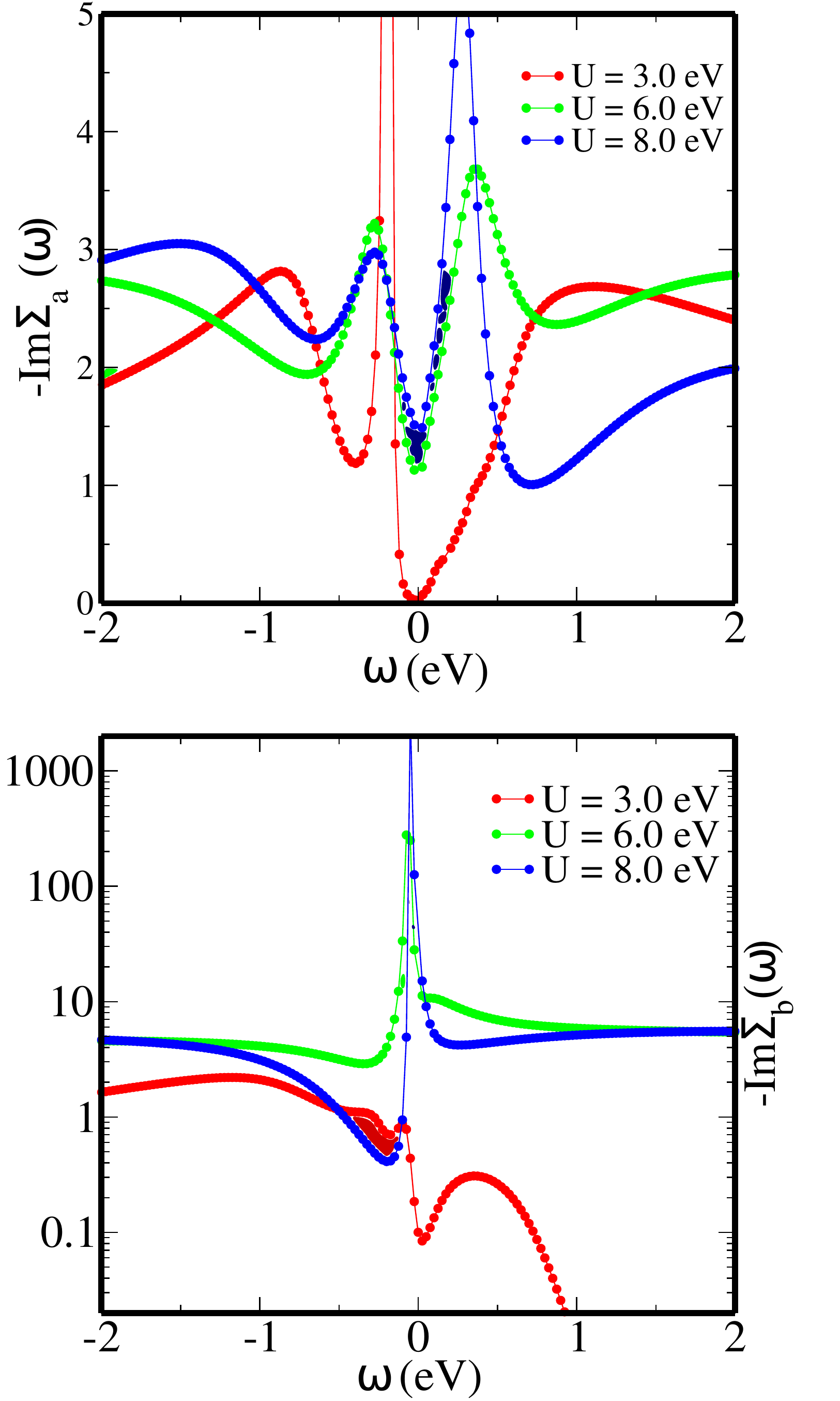}
\caption{The low energy behavior of the analytically continued -Im$\Sigma_{a}(\omega)$ and  -Im$\Sigma_{b}(\omega)$ shows the emergence of orbital selective features in the single particle sector. The large finite intercept of -Im$\Sigma_{a}(\omega)$ in the critical regime testifies for the OSMP with finite and large non-Fermi liquid scattering while -Im$\Sigma_{b}(\omega)$ shows pole structure.}
\label{self}
\end{figure}

First, we perform analytic continuation of Matsubara frequency data using maximum entropy method (MEM)~\cite{mark} to obtain real 
frequency spectra of interest. The local density of states (LDOS) are shown in Fig.\ref{dos} at $58$~K and for a range of interaction strength 
$U$ varying between $1$~eV and $11$~eV.  A strongly renormalized low-energy Landau quasi-particle at $E_{F}(=0)$ for $U$ in the range between $2.5$~eV and $4$~eV
smoothly crosses over to an incoherent continuum response in the orbital selective Mott phase for $U > U_{c,OSMT} = 5.9 eV$.  Finally, for $U > 11$~eV, a full Mott insulator obtains. In the 
OSMP, $b$ orbital-states are Mott insulating, while $a$-orbital states are metallic, but the $a$-fermion propagator, $G_{a}(\omega)$, exhibits a low-energy incoherent continuum (branch-cut) structure.
 The analytically 
continued -Im$\Sigma_{a,b}(\omega)$ (Fig.~\ref{self}) shows a pole structure for the $b$ orbital, while -Im$\Sigma_{a}(\omega)$ has a large 
but finite intercept at $\omega=0$. At lower $U(<U_{c,OSMT})$, both, -Im$\Sigma_{a,b}(\omega)\simeq \omega^{2}$ show correlated LFL behavior.  

\vspace{0.2em}
\begin{figure}[h!]
\centering
\includegraphics[angle=0,width=0.8\columnwidth]{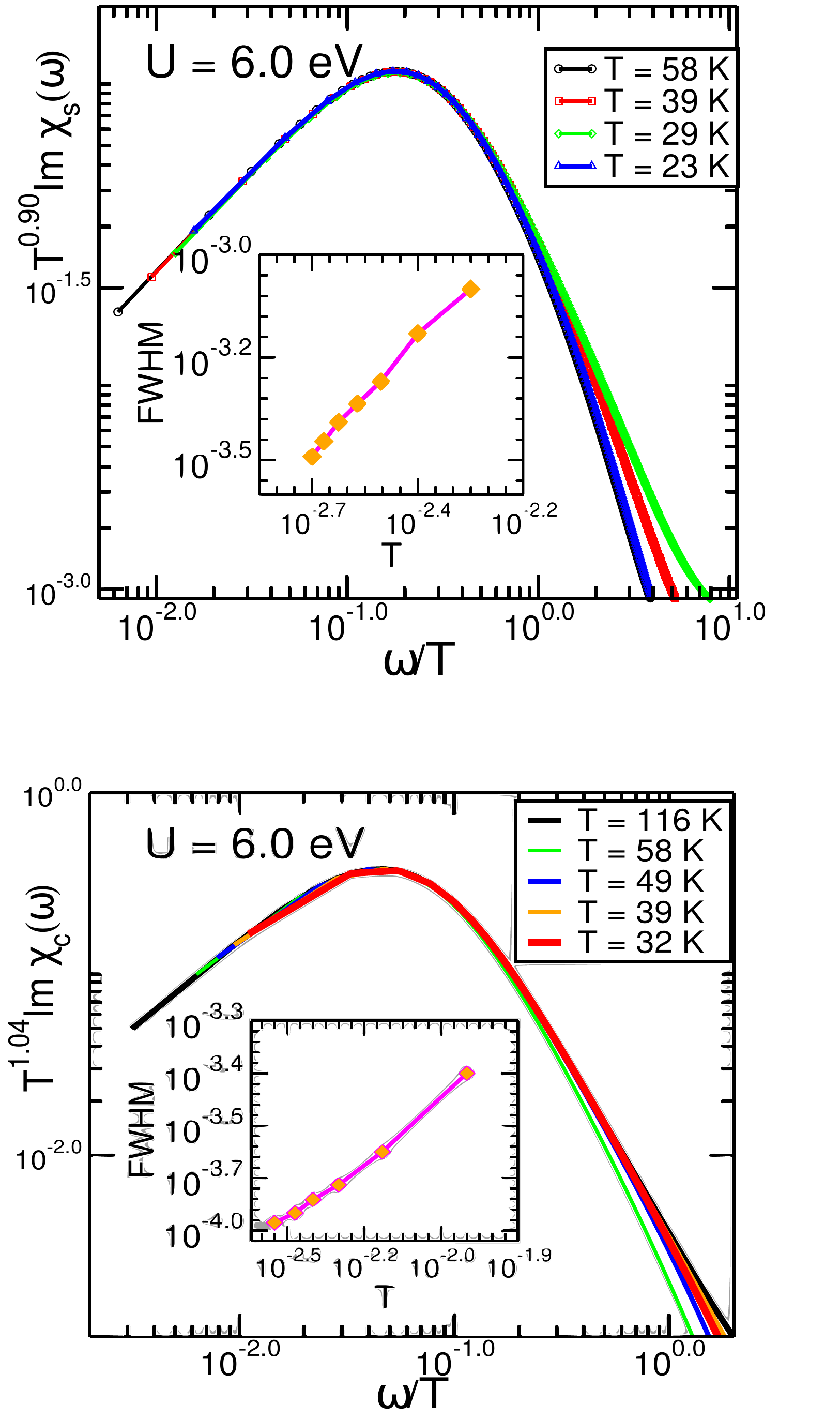}
\caption{ Im$\chi_{s}(\omega,T)$ and Im$\chi_{c}(\omega,T)$ show a proper thermal scaling collapse Im$\chi_{s}(\omega,T)\sim F(\omega/T)$ in the critical regime at $U$ = 6.0 eV. The inset in each figure shows the FWHM extracted from Im$\chi_{s}(\omega,T)/\omega$ and Im$\chi_{c}(\omega,T)/\omega$ respectively which are nearly linear over a range of temperature where the scaling collapse is perfect.}
\label{spinscale}
\end{figure}
\vspace{0.2em}

 Remarkably, in the OSMP, the local part of the dynamical spin susceptibility in Fig.\ref{spinscale} exhibits infra-red singular and 
fractional power-law scaling behavior characteristic of the strange metal: Im$\chi_{loc}^{zz}(\omega) \simeq T^{-\alpha_{s}}f(\omega/T)$ with
$\alpha_{s}\simeq 0.9$.  This clearly shows that the extinction of Landau quasiparticles in $G_{a}(\omega), G_{b}(k,\omega)$ directly manifests 
in the emergence of a critical branch-cut continuum in (single-spin-flip) spin-fluctuations.  Even more interestingly, the dynamical charge 
susceptibility as shown in Fig.~\ref{spinscale} also exhibits similar scaling form, but with an exponent $\alpha_{c} \neq 
\alpha_{s}$, implying that spin and charge fluctuations propagate with distinct velocities (See SI) - a kind of disparate spin and charge fluctuation responses 
associated with the OSMT.  Such disparate spin and charge responses are also observed in unconventional superconductors, emerging from the interplay of orbital selectivity~\cite{swagata} and spin-orbit coupling~\cite{nsrep}. We show how this unusual emergent feature results in very good accord with the unusual normal-state magnetic 
fluctuation spectra and optical conductivity in the ``strange metallic'' region in cuprates and Fe-arsenides.
 
\vspace{0.3cm}

\begin{figure}[h!]
\centering
\includegraphics[angle=0,width=0.8\columnwidth]{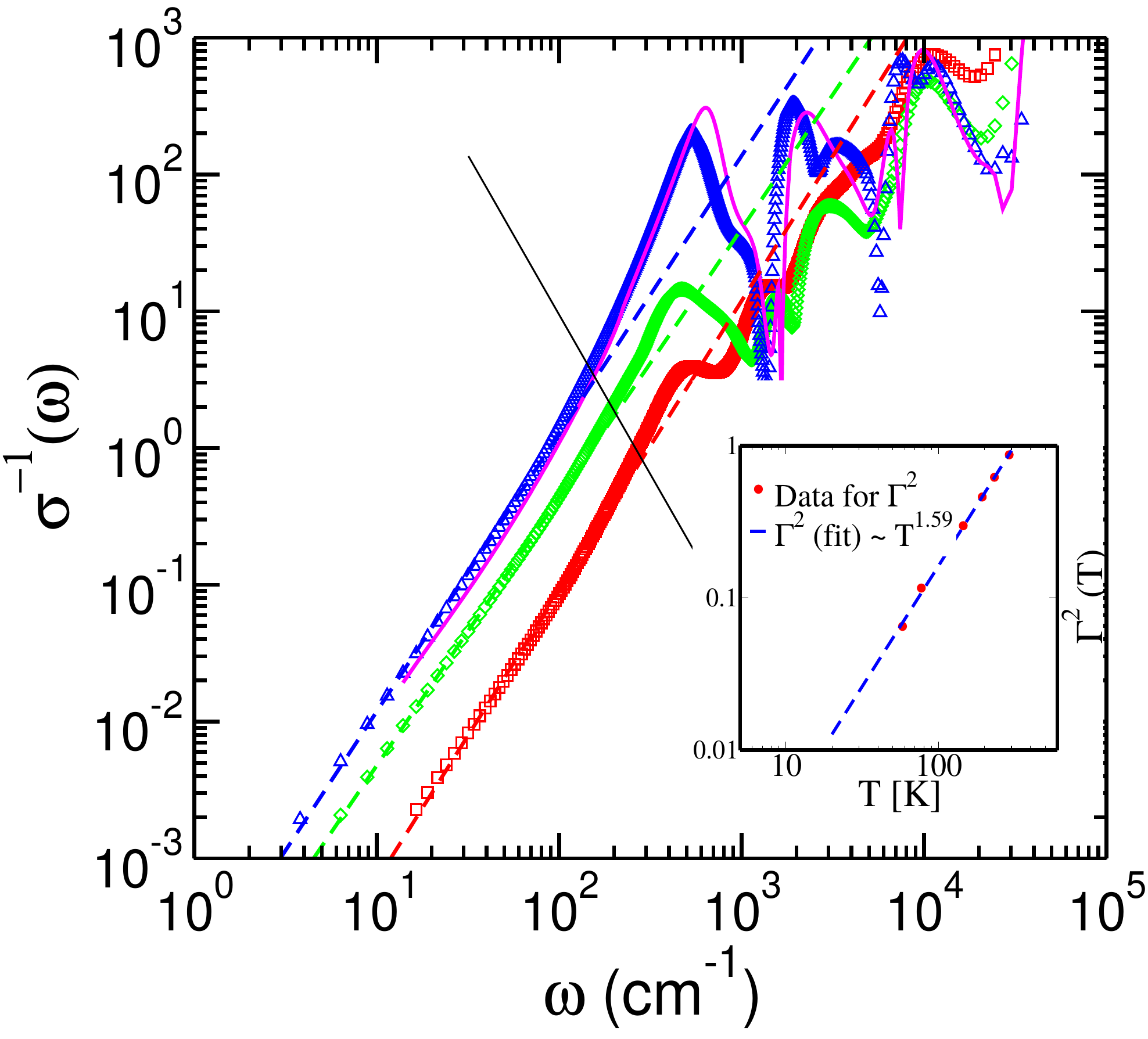}
	\caption{The inverse of $\sigma^{-1}(\omega)$ shows a quadratic frequency dependence over a range temperatures between $116$~K and $58$~K upto a certain finite energy. The critical energy upto which $\sigma^{-1}(\omega) \sim \omega^{2}$ increases with lowering temperatures. (inset) FWHM for the $\sigma(\omega, T)$ and the thermal exponent for the fitting function.}
	\label{optics}
\end{figure}

\begin{figure}[h!]
\centering
\includegraphics[angle=0,width=0.98\columnwidth]{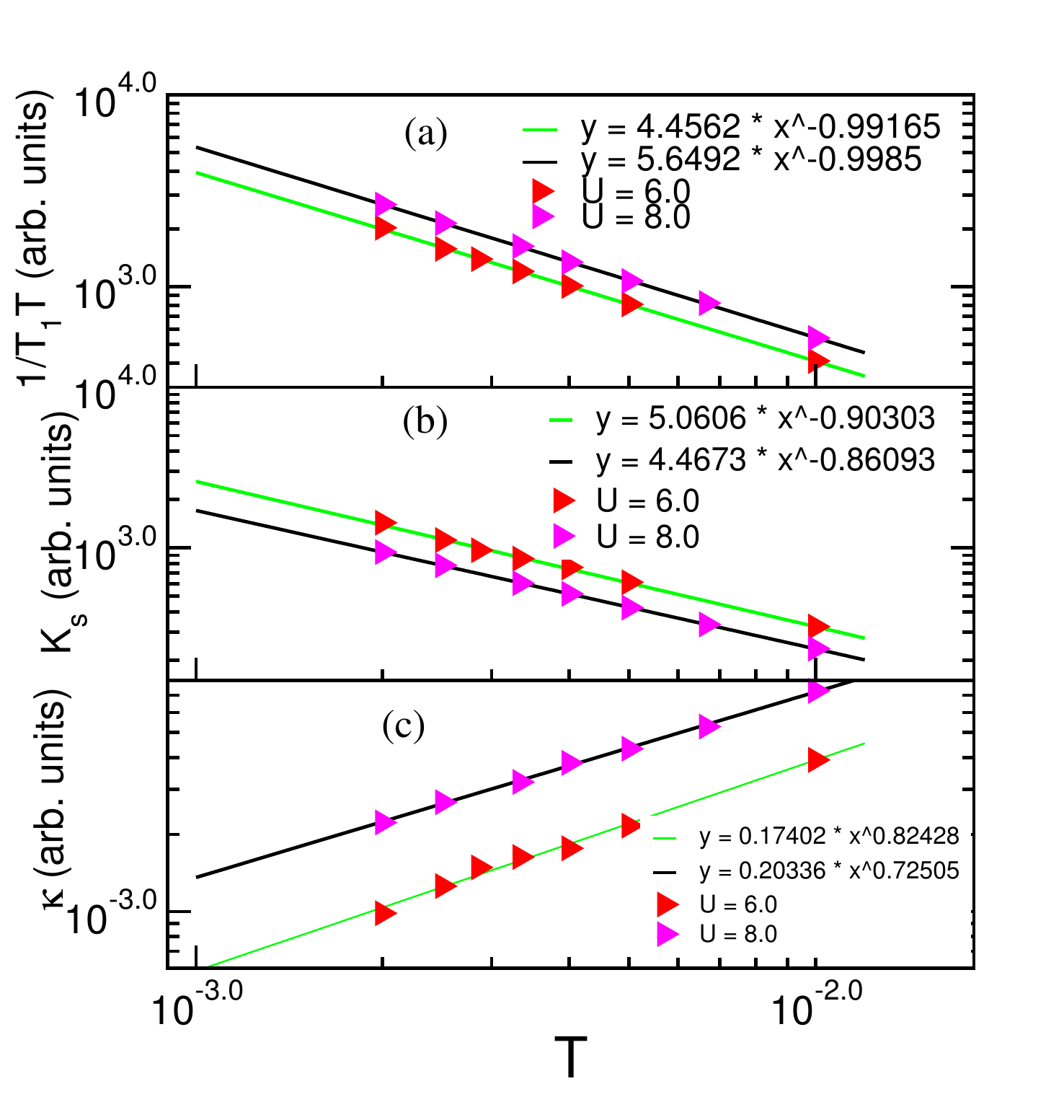}
\caption{Panel (a): the nuclear magnetic relaxation rate divided by $T$, $1/T_{1}T\simeq T^{-0.99}$, implying almost $T$-independent $1/T_{1}$,
	in good accord with Aeppli {\it et al.}~\cite{aeppli}.  Panel (b): Knight shift, $K_{s}(T)\simeq T^{-n}$ with $n=0.86,0.9$, implying breakdown of the Korringa relation in strange metals.
	Panel (c): the full-width at half-maximum (FWHM) of the spin fluctuation lineshape, $\kappa_{s}(T)\simeq a(U)+b(U)T^{m}$ with $m \simeq 0.725, 0.82$, showing the anomalous nature of spin fluctuations.}
\label{NMR}
\end{figure}

Physically, these features emerge as a direct manifestation of emergent, critical pseudoparticles driving the extinction of stable
Landau-damped FL-like collective modes in the strange metal. This is because charge  and spin fluctuations are themselves constructed from the
now incoherent continuum of one-fermion excitations, rather than usual Landau quasiparticles.  The underlying physical origin of these emergent
anomalous features is quite revealing. We first observe that in the critical metal, the tendency of $V_{ab}$ to transfer an $a$-fermion into a 
$b$-fermion is dynamically blocked in the OSMP.  This is because the lower-Hubbard 
band now corresponds to singly occupied $b$-states, so action of $V_{ab}$ must create a doubly occupied  
(two opposite-spin electrons in the $b$-orbital) intermediate state.  However, this lies in the upper
Hubbard band in the $b$-sector and thus the resulting term now has the form 
$V_{ab}'(n_{i,b,-\sigma}b_{i,\sigma}^{\dag}a_{j,\sigma}+h.c)$, which couples the $a$-fermion to a upper-Hubbard band $b$-fermion,
and thus has no interpretation in terms of a coherent one-electron-like state any more.  
In this sector, this is a high-energy state, and so $V_{ab}(k)$ leads to a ``UV-IR'' mixing between low- and high-energy states. It is this emergent projective dynamics that is at the root of irrelevance of $V_{ab}(k)$ at 
one-electron level~\cite{pwa} and emergence of 
strange metal features we find above.  An interesting aspect of our results is that the branch cut continuum features in two-particle sector arise due to the interband $U_{ab}$
for $U_{ab}>U_{ab}^{c,OSMT}$: this implicates strong, coupled charge- (valence in multi-orbital Anderson case) and spin fluctuations in destruction of LFL metallicity.

At two-particle level a divergent number of soft, local spin fluctuation modes, manifests itself 
as an infra-red singularity in the transverse spin fluctuation spectrum (in fact, such a singularity in the ``spin-flip excitonic'' correlator is expected in
an X-ray edge problem), along with local quantum critical $\omega/T$ scaling and anomalous exponents in the fluctuation spectra. 


  Remarkably, it turns out that our results afford a consistent quantitative description of both 
optical and inelastic neutron results in the strange metal.  In Fig.~\ref{NMR}, we show the NMR 
$1/T_{1}T=A$ Lim$_{\omega\rightarrow 0}\frac{\chi_{s}''(\omega)}{\omega}$, the Knight shift 
$K_{s}=B.$Re$\chi_{s}(q=0,T)=B.\chi_{loc}(T)$ ($A,\, B$ are constants) and the spin fluctuation damping $\kappa$, defined as 
full-width at half-maximum in Im$\chi_{s}({\bf q},\omega,T)$.  First, we find $1/T_{1}T\simeq 1/T^{x}$ with $x\simeq 1$, implying 
$1/T_{1}=constant$.  But the Knight shift $K_{s}(T)\simeq T^{-n}$ with $n=0.9\, (U=6.0)$ and $n=0.86\, (U=8.0)$, and thus the Korringa 
relation $1/[T_{1}TK_{s}^{2}]=constant$ is {\it not} satisfied.  Concomitantly, the spin fluctuation linewidth, 
$\kappa_{s}(T)\simeq c + bT^{m}$ with $m=0.82 \, (U=6.0)$ and $m=0.72 \, (U=8.0)$.  Taken together, $1/T_{1}$ and $K_{s}(T)$ along
with the $\omega/T$-scaling in $\chi({\bf q},\omega)$ are in very good accord with data in the strange metal region of the cuprate phase diagram.  
To study the charge dynamics, we have used the analytically continued one-particle DMFT Green functions along with the Kubo formalism 
(whence the irreducible vertices drop out in the Bethe-Salpeter equations for conductivity) to compute the ac longitudinal conductivity, 
$\sigma_{xx}(\omega)$ in Fig.~\ref{optics}.  Direct comparison with data~\cite{akrap} shows surprisingly good accord with data, over an 
extended energy range.  Specifically, our main finding is the anomalous power-law infra-red response: $\sigma_{xx}(\omega)\simeq \omega^{-1.2}$ 
up to 300 meV, very close to the data, crossing over to a ``relaxational dynamics'' form $\sigma_{xx}(\omega,T)\simeq (\omega^{2}+\Gamma^{2}(T))^{-1}$,
but with an anomalously damped $\Gamma(T)\simeq T^{0.8}$ (see inset of Fig.~\ref{optics}) at lower energy.  
More interestingly, the mid-infrared peak around $\Omega\simeq 0.1$~eV is also closely reproduced in our results in very good accord with data.   
These observations strongly support a quantum critical {\it phase} for $U_{ab}>U_{ab}^{(c)}$ beyond which an OSMP obtains.

  Comprehensive description of both charge and spin fluctuations in the strange metal is thus 
found to be intimately related to an (orbital) selective
Mott phase at intermediate-to-strong coupling in the two-band Hubbard model.  As long as an 
OSMP occurs in DMFT, blocking of $b$-fermion recoil during scattering (cf. the 
$U_{ab}\sum_{i}n_{ia}n_{ib}$) wipes out the lattice 
(Landau-FL) coherence scale via Kondo breakdown and associated X-ray-edge anomalies, 
and results in vanishing Landau quasiparticle residue, $z_{LFL}(\omega=0)=0$.  Loss of the 
OSMP at smaller $U$ goes hand-in-hand with relevance of $V_{ab}(k)$ at the one-electron 
level, cutting off the X-ray edge singularities and restoring correlated LFL behavior.  
These findings are further bolstered by an analytic bosonization approach (see SI) to the impurity 
problem wherein, following Anderson~\cite{pwa}, the inverse orthogonality catastrophe in 
the impurity problem of DMFT is argued to lead to anomalous power-laws with $\omega/T$-scaling
in the fluctuation propagators. This accords with our CTQMC results, providing a microscopic interpretation of our numerics.  
In our two-band model, the $d$-wave form factor in $V_{ab}(k)$ aids in the development of 
nodal-antinodal (N-AN) differentiation, connecting our findings to anomalous responses in 
nodal metals.  Indeed, Homes {\it et al}~\cite{akrap} have linked the 
anomalous $\sigma_{xx}(\omega)$ to nodal metallicity.  Crucially, however, we find that an 
OSMP and the resultant branch-cut structure of the one- and two-particle propagators are 
{\it necessary} to describe the specific strange metallic anomalies.  Finally, a topological 
Fermi surface reconstruction should generically accompany the transition between a correlated 
LFL and an OSMP, since the $b$-fermion FS is extinguished by selective-Mottness, while the $a$-fermion FS
still exists, since Im$\Sigma_{a}(\omega=0)=0$ in spite of extinction of the Landau quasiparticle pole: this will result in a transition from a large to a small Fermi surface, which is also frequently observed across QPTs characterized by strange metallicity.

To summarize, we present strong numerical evidence linking the famed strange metal 
anomalies in cuprates and other correlated systems to onset of an orbital-selective Mott 
phase (OSMP), characterized by {\it orbital freezing}, in a multi-band Hubbard model.  Following Anderson and Casey~\cite{pwa}, a deeper investigation into the structure of 
charge and spin correlations upon bosonization of the impurity model (see SI) reveals 
interesting connections with high-dimensional spin-charge separation, providing important insight 
into our CTQMC results.  This link rationalizes the good accord with a range of charge and spin 
fluctuation responses in the strange metal: our results are applicable to cuprates if we 
invoke the ``cluster-to-orbital'' mapping~\cite{werner}, where an embedded 4-site cluster was mapped onto a ``two-orbital'' Hubbard model using the cluster-to-orbital mapping.  Then, thanks to this mapping, the OSMP turns out to be the momentum-selective Mott phase in a new guise.  In its multi-orbital form, our model and results should also apply to multiband systems like Fe-arsenides and $f$-electron systems in
appropriate parameter regimes of our two-band model.  How such a critical multi-electronic 
continuum as we find is implicated in generating residual interactions that lead to 
unconventional superconductivity (USC) and competing orders, however, remains an outstanding open issue.    

SA acknowledges Simons Many-Electron Collaboration and UGC (India) for research fellowships. Additional support (MJ) was provided by NSF Materials Theory grant DMR1728457.

\bibliographystyle{apsrev4-1}

\section{Supplemental Material}
\begin{figure}[h!]
\centering
\subfigure[]{\label{f:C11}\epsfig{file=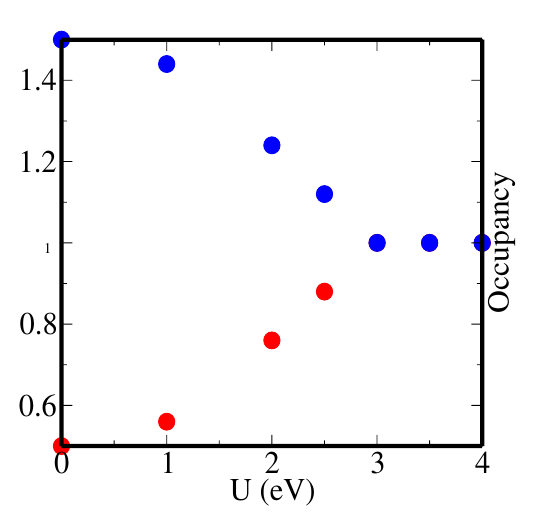,trim=0in 0in 0in 0.0in,
clip=true,width=0.49\linewidth}}\hspace{-0.0\linewidth}
\subfigure[]{\label{f:C11}\epsfig{file=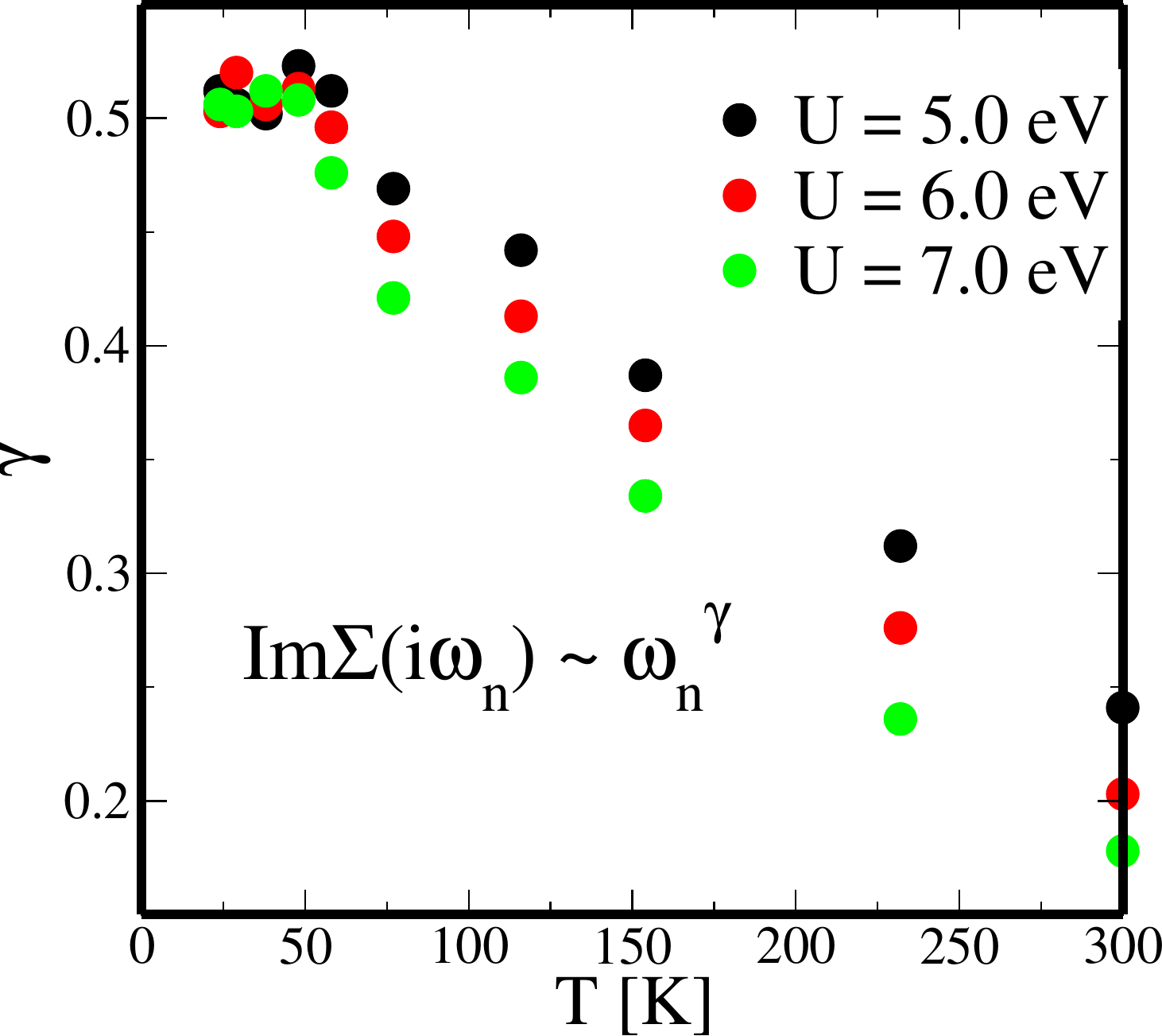,trim=0in 0in 0in 0.0in,
clip=true,width=0.49\linewidth}}\hspace{-0.0\linewidth}
\subfigure[]{\label{f:C21}\epsfig{file=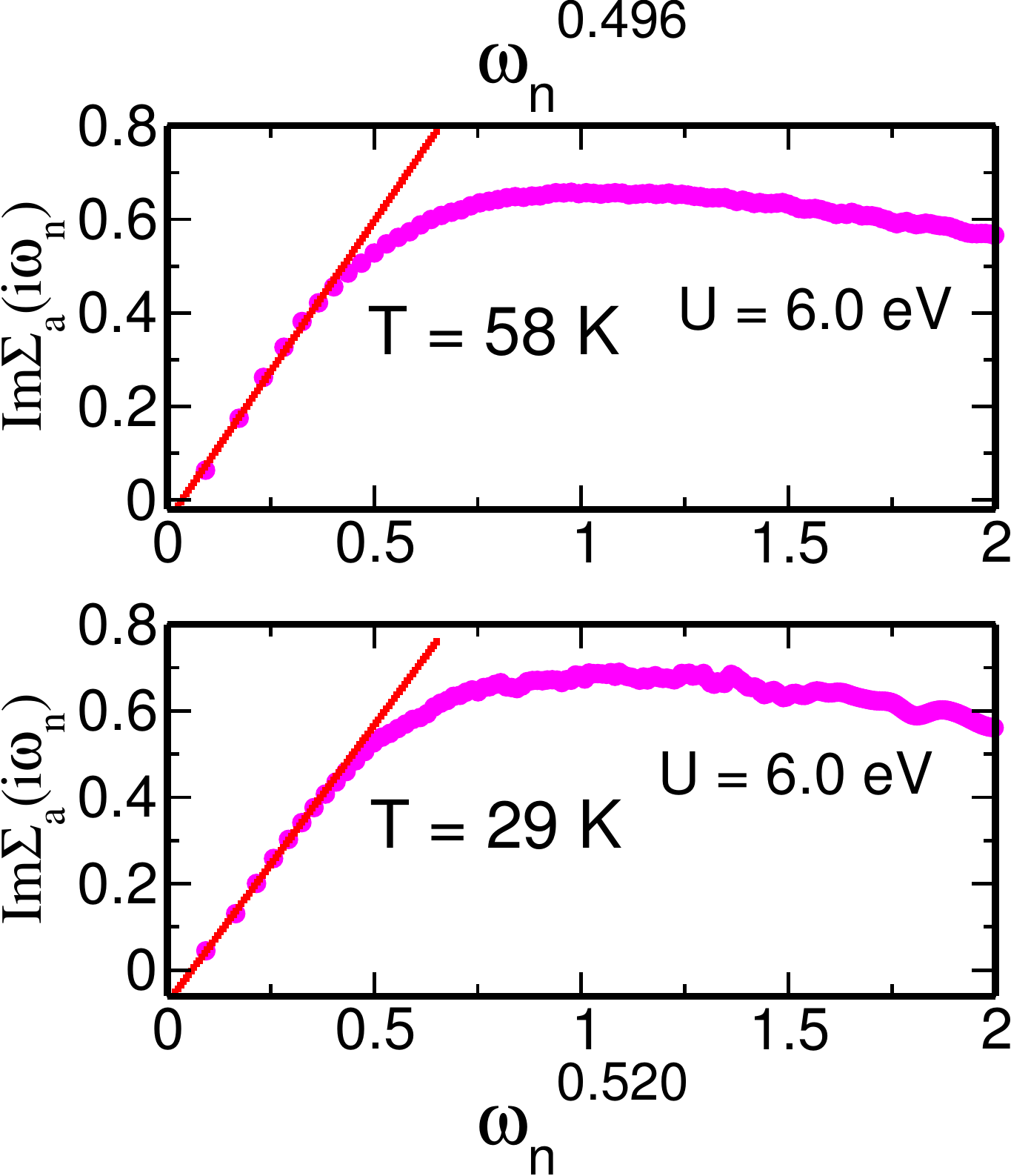,trim=0in 0in 0in 0.0in,
clip=true,width=0.49\linewidth}}\hspace{-0.0\linewidth}
\caption{(a) The evolution of the occupancies for each orbital is shown. The local U(1) emerges as U is cranked up. -$Im\Sigma(\omega_{n})$ is fitted to a form -$Im\Sigma(\omega_{n})= C + A (\omega_{n})^{\gamma}$.
 In the fermi liquid regime $\gamma$ must approach $1.0$. In the critical regime, the exponent is robust against interaction 
 parameters. The robustness of such critical behavior of single particle self-energy is shown (b). The low energy power law fit to the 
 self energy (c) shows that the exponent evolves with temperatures before finally saturating to $\sim 0.5$ at lowest temperatures.}
\label{expoself}
\end{figure}

 We show in Fig.~\ref{expoself} how the occupancies of individual orbitals evolve with $U$. Beyond $U = 3 eV$ both the orbitals become half-filled, whence an {\it emergent} particle-hole symmetry emerges. We show the imaginary 
part of the self-energies, Im$\Sigma_{a}(i\omega_{n})$ for a range of correlation parameters. The low energy features of 
the imaginary part of self-energy is fitted to a form $C + A\omega_{n}^{\gamma}$. The exponent $\gamma$ as shown in Fig.~\ref{expoself} in the 
OSMT phase shows a robust behavior for range of $U$, more specifically at lower temperatures, where it saturates to a value of $0.5$.  This is clear evidence for 
non-LFL metallicity in the OSMP.  

\vspace{0.5em}

We stress the robustness of the critical thermal scaling collapse for Im$\chi_{s}(\omega,T)$. Here, we show it for a different $U > U_{c,OSMT}$
which again shows an excellent thermal scaling collapse for $T^{-\alpha_{s}}$Im$\chi_{s}(\omega,T)\sim F(\omega/T)$ at $U$ = 8.0 eV 
(Fig.~\ref{spinscale}) with a slightly different exponent. This shows that the anomalous scaling persists over a range of $U/t_{a,b}$, attesting to
its robustness, and indicates a quantum critical {\it phase}.  Moreover, the full-width at half-maximum (FWHM) extracted from Im$\chi_{s}(\omega,T)/\omega$ is linear in $\omega$ at low $T$.
\vspace{0.5em}
\begin{figure}[h!]
\centering
\subfigure[]{\label{f:C21}\epsfig{file=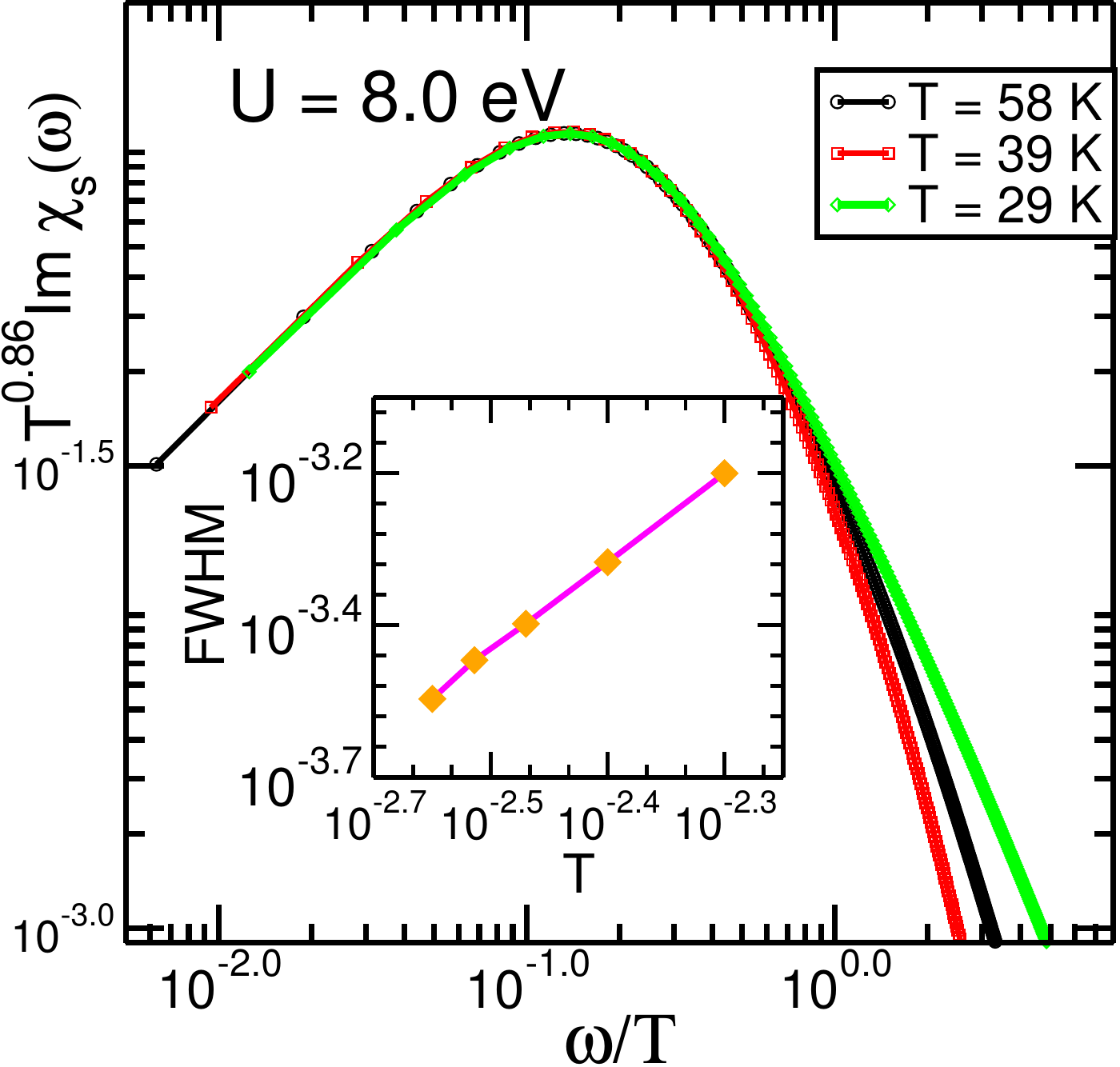,trim=0in 0in 0in 0.0in,
clip=true,width=0.98\linewidth}}\hspace{-0.0\linewidth}
\caption{ Im$\chi_{s}(\omega,T)$ shows a proper thermal scaling collapse Im$\chi_{s}(\omega,T)\sim F(\omega/T)$ in the critical regime at $U$ = 8.0 eV. The inset shows the FWHM extracted from Im$\chi_{s}(\omega,T)/\omega$ which is nearly linear over a range of temperature where the scaling collapse is perfect.}
\label{spinscale}
\end{figure}
\vspace{0.5em}

   Finally, the fact that the one- and two-fermion  propagators obey $\omega/T$-scaling in the OSMP with fractional exponents is interesting. First, this implies that the corresponding relaxation rates, evaluated from $\Gamma_{M}(T)=i[\partial$ln$ M(\omega,T)/\partial\omega]|^{-1}_{\omega=0}$
with $M=G_{aa}(\omega),\chi_{s}^{loc},\chi_{c}^{loc}$, are all {\it linear} in $T$.  That the exponents in the power-law behavior in spin- and charge fluctuation propagators is distinct reflects the importance of vertex corrections.  These are absent in both IPT and large-$N$ solvers in the DMFT context, but are encoded in CT-QMC.  Second, our finding that relaxation rates are linear in $T$ also shows the interacting character
(noticed earlier by Glossop {\it et al.}~\cite{glossop} of the ``strange'' metal phase, wherein non-linear coupling between quantum critical modes is finite.  Thus, our finding of distinct exponents in the dynamical spin- and charge fluctuations is {\it not} an artifact.  Experimentally, such behaviour has been seen in $f$-electron systems near local quantum-critical {\it points}(see Glossop {\it et al.} ~\cite{glossop} and references therein).

{\bf Analytic rationalization}

\vspace{0.3cm}

Here, we describe how our CTQMC results can be understood analytically.  
As in hidden-FL or FL$^{*}$ theories, we will not need to invoke proximity 
to a $T=0$ antiferromagnetic ordered state to rationalize our findings.
This is because DMFT accesses the dynamical but {\it local} spin and charge fluctuations
and the changes in their analytic structure as the OSMP is approached from the LFL side
as $U_{aa,bb},U_{ab}$ are increased.
In contrast, in the spirit of Anderson and Casey~\cite{casey}, we show that using the Schotte-Schotte approach to bosonization of the impurity model 
of DMFT provides a clean interpretation of our results in terms of bosonic ``tomonagons''.

  {\bf Analytic Insight:}  
  
  Our results bear similarity to the hidden-FL view of Anderson as follows.  In the hidden-FL theory, applied deep in the doped Mott
  insulator phase of a $t-J$ model, the exact eigenstates $|\Psi\rangle$ are related to the unprojected states $|\Phi\rangle$ by a Gutzwiller
  projection: $|\Psi\rangle =\Pi_{i}(1-n_{i\uparrow}n_{i\downarrow})|\Phi\rangle=P|\Phi\rangle$.  Using $Pc_{i\sigma}^{\dag}P=c_{i\sigma}^{\dag}(1-n_{i,-\sigma})$,
  the one-electron Green function $G_{ij}(t)=-i\langle\Psi|T[c_{i\sigma}(t)c_{j\sigma}^{\dag}(0)]|\Psi\rangle$ is written (in the $U\rightarrow\infty$ limit)
  as $G_{ij}(t)=G_{ij}^{0}(t)G_{ij}^{*}(t)$, where $G_{ij}^{0}(t)=-i\langle\Phi|c_{i\sigma}(t)c_{j\sigma}^{\dag}(0)|\Phi\rangle$ is the free electron propagator, 
  and $G_{ij}^{*}(t)=\langle\Phi|T[(i-n_{i,-\sigma})(t)(1-n_{j,-\sigma})(0)|\Phi\rangle$ represents the scattering processes involving opposite
  spin fermions due to the Hubbard interaction.  Arguing that the latter propagator can be computed by making analogy with the seminal ``X-ray edge'' problem,
  Anderson concludes that $G_{ij}(t)\simeq t^{-(1+\eta)}$ with $\eta$ an ``s-wave'' scattering phase shift at the Fermi surface.  The fact that 
  $0< \eta <1$ implies that the infra-red pole structure of $G(k,\omega)$ is replaced by a branch-cut singularity, leading to non-LFL metallicity. 
  In our model (see main text), onset of the OSMP gives rise to a similar projective aspect due to Mott localization of $b$-states: in the OSMP, at low energy below the 
  selective-Mott gap in $G_{bb}(\omega)$, the $b$ states cannot recoil during an $a-b$ fermion scattering process (which transfers an $a$-fermion into a $b$-fermion state), simply because there are now
  {\it no} lower-Hubbard band states into which they can recoil.  $V_{ab}(k)$ can now only create {\it upper}-Hubbard band states in the $b$-fermion sector, as described in the text.
  The above argument of Anderson can now be applied to the non-local hybridization (or inter-band hopping) term in identically the same way as above, with 
  precisely the same result: Landau FL metallicity as signified by an infra-red pole structure of $G_{aa}(\omega)$ is destroyed due to the emergent projective aspect in the $b$-fermion sector which 
  controls the low-energy physics in the OSMP.

\section{Bosonization}

   In this section, we detail how the projective aspect at the root of emergence of ``strange'' metallicity permits further analytic insight.  Specifically, it allows us to use bosonization for the underlying impurity model in the OSMP regime to analytically rationalize the high-dimensional spin-charge separation, wherein the exponents characterizing the power-law decay of dynamic spin and charge correlations are distinct.  We now analyze a suitable impurity limit of the two-band Hubbard Model (2BHM) via bosonization to analytically 
rationalize this exciting feature observed (main text) in the DMFT calculations using CT-QMC.  We emphasize that it is important to notice that, in the orbital selective Mott phase (OSMP) 
of the 2BHM, the Mott-localized $b$-orbital states interact with the metallic 
$a$-orbital states via U$_{ab}$ and J$_{H}$ in the regime where the 
interband one-electron hybridization, $V_{ab}(k)$, is irrelevant: as detailed in the main text, this is the
regime where DMFT for the two-band Hubbard model yields an OSMP. 

This impurity model is then written as 

\begin{align*}
H_{imp} & = \sum_{k,\sigma}\epsilon^{\sim}_{ka}a^{\dagger}_{k,\sigma}a_{k,\sigma}+Un_{oa\uparrow}n_{0a\downarrow}+ U_{ab}n_{oa}n_{ob} \\
          & -J{\bf S}_{oa}.{\bf S}_{ob}-\mu\sum_{\sigma} n_{oa\sigma}
\end{align*}

where $^{\sim}\epsilon_{ka}$ is the $a$-band dispersion in the OSMP, and the $b$-band states are understood to be the Mott localized , i,e., the lower $b$-Hubbard band consists of single occupied states and double occupancy of 
$b$-band states is forbidden in the asymptotic low-energy limit. 

To clarify the roles of $U_{ab},J$ in the emergence of the novel features, we begin with U$_{ab}$=0 and J = 0, where low-energy correlated Landau FL metallicity obtains, and consider their effects later. The impurity model
\begin{equation}
H^{0}_{imp}=\sum_{k,\sigma}\epsilon^{\sim}_{ka}a^{\dagger}_{k,\sigma}a_{k,\sigma}+Un_{oa\uparrow}n_{oa\downarrow}-\mu\sum_{\sigma} n_{oa\sigma}
\end{equation}
can be recast as,
 
\begin{align*}
H^{0}_{imp}& =\sum_{\sigma}[iv_{F}\int_{-\infty}^{\infty}]dx\psi_{\sigma}^{\dagger}(x)\delta_{x}\psi_{\sigma}(x)+\frac{U}{2}:\psi_{\sigma}^{\dagger}(0) \\
            & \psi_{\sigma}(0)::\psi_{-\sigma}^{\dagger}(0)\psi_{-\sigma}(0):-\mu:\psi_{\sigma}^{\dagger}(0)\psi_{\sigma}(0):
\end{align*}

where $\psi_{\sigma}(x)$ are chiral (right-moving) fermion fields 
describing the radial (outward and inward from the impurity, ``o'') band motion of $a$-fermions.
v$_{f}$=$\delta_{k}\epsilon_{ka}$|$_{k=k_{F}}$, $:A:$ implies normal ordering of A, i.e., 
$:A:$ = A - $\langle 0|A|0\rangle$ with $|0\rangle$ denoting the ground state. 

Next, use the bosonization identity

$\psi_{\sigma}(x)=\frac{1}{\sqrt{2\pi\alpha}}e^{i\phi_{\sigma}(x)}, \phi_{\sigma}(x)=\sqrt{\pi}[\phi_{\sigma}(x)-\int_{-\infty}^{x}dx^{'}\Pi_{\sigma}(x^{'})]$

where $\psi_{\sigma}(x)$, $\pi_{\sigma}(x)$are conjugate bosonic fields satisfying
 
$[\phi_{\sigma}(x),\Pi_{\sigma'}(x')]=i\delta_{\sigma\sigma'}\delta(x-x'),$ and $\alpha$ is a short-distance cut-off.

Introducing the charge and spin-fields $\phi_{c}$=$\sum_{\sigma}\phi_{\sigma}$, 
$\phi_{c}$=$\sum_{\sigma}\sigma\phi_{\sigma}$, we can ``split'' H$^{0}_{imp}$ into 
charge (c) and spin (s) sectors as  H$^{0}_{imp}$ = H$^{0,c}_{imp}$ + H$^{0,s}_{imp}$:

\begin{align*}
H_{imp}^{0,c} & = \frac{v_{F}}{2}\int_{-\infty}^{\infty}dx[\Pi_{s}^{2}(x)+(\delta_{x}\phi_{s}(x))^{2}] \\
               & -\frac{U}{\sqrt{2}\pi}(\delta_{x}\phi_{s}(0))^{2}-\frac{U}{8\pi^{2}}(\delta_{x}\phi_{s}(0))^{2}
\end{align*}

$:\psi_{\sigma}^{dagger}(0)\psi_{\sigma}(0):=\frac{1}{2\pi}\delta_{x}\phi_{\sigma}(0)$, and $n_{ob\downarrow}=\langle\psi_{\downarrow}^{\dagger}(0)\psi_{\downarrow}(0)$

Now we resolve the bosonic field operates into their Fourier components as,

\begin{equation}
\phi_{\nu}(x)= \sum_{k}\frac{1}{\sqrt{2|k|}}(a_{\nu,k}e^{ik_{x}})+a^{\dagger}_{\nu,k}e^{-ik_{x}})e^{-\alpha|k|/2}
\end{equation}

\begin{equation}
\Pi_{\nu}(x)= -i\sum_{k}\sqrt{|k|/2}(a_{\nu,k}e^{ik_{x}})-a^{\dagger}_{\nu,k}e^{-ik_{x}})e^{-\alpha|k|/2}
\end{equation}

%

with $\nu$=c,s. Eqns. (5),(6) become

\begin{align*}
H^{o,c}_{imp}& =\sum_{k\rangle 0} \omega_{k} {a_{ck}}^{\dagger}a_{ck}+i{\sqrt{2\rho}}(Un_{ob\downarrow}-\mu)\sum_{k\rangle 0}\sqrt{\omega_{k}}(a_{ck}\\&-{a_{ck}}^{\dagger})-\rho\frac{U}{2}\sum_{k,k'\rangle 0}(a_{ck}-a_{ck}^{\dagger})(a_{ck'}-a_{ck'}^{\dagger})
\end{align*}

\begin{equation}
H^{o,s}_{imp}=\sum_{k\rangle 0} \omega_{k}a_{sk}^{\dagger}a_{sk}+\rho\frac{U}{2}\sum_{k,k'\rangle 0}\sqrt{\omega_{k}\omega_{k'}}(a_{sk}-a_{sk}^{\dagger})(a_{sk'}-a_{sk'}^{\dagger})
\end{equation}

where $\omega_{k}$ = kv$_{F}$ and $\rho$ = $\frac{1}{2\pi v_{F}}$

A this point, one can show that correlated Landau FL properties follow 
upon using an equation of motion approach.
In particular, 
the local charge susceptibility,

$\chi_{c}$ = $\frac{2}{\rho(U+U_{c})}$ + $\frac{i\omega}{2\pi v_{F}^{2}} \frac{2U_{c}^{2}}{(U+U_{c})^{2}}$

The local spin susceptibility,

$\chi_{s}$ = $\frac{1}{(U_{c}-U)}$ + $\frac{i\omega}{2\pi v_{F}^{2}} \frac{U_{c}^{2}}{(U-U_{c})^{2}}$

The LFL behavior is clear and as expected, $\chi_{c}$ is suppressed while $\chi_{s}$is enhanced with increasing U.

\subsubsection{Effects of U$_{ab}$ and J}

In the OSMP, the metallic $b$-fermions strongly scatter off the selectively Mott-localized $a$-fermions via both, $U_{ab}$ and $J$.  We will now show how the different exponents in the power-law fall-off of the charge and spin fluctuation propagators in the infra-red, found in DMFT(CTQMC) studies in the main text, 
emerge as a consequence of $(i)$ the mapping of the underlying impurity problem onto the famed X-ray edge problem, and $(ii)$ the different scattering potential experienced by ``metallic'' $a$-fermions in the charge and spin-fluctuation channels.
 
First, consider the effect U$_{ab}$. Since the a-states are Mott localized, 
$H_{U_{ab}}$=U$_{ab}$n$_{oa}$n$_{ob}$ in Eq.(1) describes the scattering of 
metallic $a$-fermions off Mott-localized $b$-fermions: importantly, due to the asymptotically valid projection of double occupancies in the $b$-fermion sector, the $b$-fermions now cannot recoil during the scattering by U$_{ab}$, simply because there are {\it no} empty lower-Hubbard band states in the $b$-fermion sector into which they can recoil 
(since the lower Hubbard band of a-sector correspond to singly occupied states).  This leads to an {\it exact} mapping of this case to the famed X-ray edge problem (PWA). In bosonized form~\cite{schotte}, 
U$_{ab}$n$_{oa}$n$_{ob}$ becomes 
U$_{ab}\sum_{\sigma,\sigma^{'}}:\psi^{\dagger}_{\sigma}(0)\psi_{\sigma}(0):n_{oa\sigma^{'}(0)}$. Thus, 
in the charge sector we get

$H^{c}_{imp} = H^{o,c}_{imp} + U_{ab}\sum_{\sigma}:\psi^{\dagger}_{\sigma}(0)\psi_{\sigma}(0):n_{oa\sigma'}(0)$

Using $:\psi^{\dagger}_{\sigma}(0)\psi_{\sigma}(0): = \frac{1}{2\pi}\delta_{x}\phi_{\sigma}$, we thus find that

 \begin{align*}
 H_{imp}^{c} & =\frac{v_{F}}{2}\int_{-\infty}^{\infty}dx[\Pi_{c}^{2}(x)+(\delta_{x}\phi_{c}(x))^{2}] \\
            & -\frac{U/2-\mu+U_{ab}n_{a}}{\sqrt{2}\pi}(\delta_{x}\phi_{c}(0))^{2}+\frac{U}{8\pi^{2}}(\delta_{x}\phi_{c}(0))^{2}
 \end{align*}

At this level (J=0), the spin sector remains unaffected. In terms of the $a_{ck},a^{\dagger}_{ck}$ oscillator modes, we now have

\begin{align*}
H^{c}_{imp} & =\sum_{k\rangle 0} \omega_{k}a_{ck}^{\dagger}a_{ck} +i\sqrt{2\rho}(Un_{ob\downarrow} \\
            & -\mu+U_{ab}n_{oa})\sum_{k\rangle 0}\sqrt{\omega_{k}}(a_{ck}-a_{ck}^{\dagger}) \\
            & -\rho\frac{U}{2}\sum_{k,k'\rangle 0}(a_{ck}-a_{ck}^{\dagger})(a_{ck'}-a_{ck'}^{\dagger})
\end{align*}

Because the one-electron hybridization is irrelevant, $U_{ab}n_{oa}n_{ob}$ incoherently scatters a- and b-fermions from the impurity into the bath and vice-versa. Thus. a propagating b-fermion 'sees' either a local potential $U_{ab}$ (when n$_{oa}$=1) or $0$ (when n$_{oa}$=0) as a function of time. Thus, the term $U_{ab}n_{oa}\sum\sqrt{\omega_{k}}(a_{ck}-a_{ck}^\dagger)$ in Eq. (13) above acts to 'shift' the charge-bosonic modes (1st term in Eq. (13)) in precisely the same way as the venerated X-ray edge problem. Following Schotte and Schotte, we can now write two Hamiltonians, corresponding to n$_{oa}$ = 0 (H$_{I}$; no scattering) and n$_{oa}$ = 1 (H$_{F}$; U$_{ab}$-scattering). Employing a unitary transformations, which is nothing but the boundary condition changing operator of Affleck et al. as

\begin{equation}
H_{F}=U^\dagger H_{I}U, U = exp[i\frac{2\delta}{\pi}\phi_{c}(0)]
\end{equation}

with $\delta=\frac{U_{ab}}{\sqrt{2}v_{F}}\frac{U_{c}}{U+U_{c}}$, the two-particle correlator,

\begin{align*}
S(t) & =\langle a_{o\sigma'}^{\dagger}(t)\psi_{\sigma}(t)\psi_{\sigma}^{\dagger}(0)a_{o\sigma'}(o)\rangle \\
     & = \langle U_{\sigma'}(t)\psi_{\sigma}(t)\psi^{\dagger}_{\sigma}(0)U_{\sigma'}(o)\rangle \sim t^{2\delta/\pi - (\frac{\delta}{\pi})^{2}}
\end{align*}

giving, 

\begin{equation}
-ImS(\omega) \sim \frac{sin[\pi(2\delta/\pi-(\delta/\pi)^{2}]}{|\omega|^{(2\delta/\pi-(\delta/\pi)^{2})}}
\end{equation}

At finite T, this diplays explicit $\omega/T$ scaling, rationalizing the DMFT qualitatively. Here, $\delta = tan^{-1} (U_{ab}\rho(0))$.
   Next, consider the term $H_{J}=J{\bf S}_{0a}.{\bf S}_{0b}$.  In a partially filled two-orbital Hubbard model in its OSMP state, the effective on-site $J$ is negative, leading to tendency to local high-spin (HS) state: however, this does not mean a tendency to ferromagnetism, since the inter-site exchange 
between local moments is usually antiferromagnetic.  Here, we focus on the quantum paramagnetic state.

 Now $H'=H_{U_{ab}}+H_{J}$ is expressible as
$(U_{ab}+J/4)\sum_{\sigma}n_{0a\sigma}n_{0b\sigma} + (U_{ab}-J/4)\sum_{\sigma}n_{0a\sigma}n_{0b,-\sigma} + J/2\sum_{\sigma}a_{0\sigma}^{\dag}a_{0,-\sigma}b_{0,-\sigma}^{\dag}b_{0\sigma}$.  Since we focus on the OSMP with Mott localized $b$-fermion states, the effect of these terms in $H'$ is {\it exactly} similar to the effect that obtains in the seminal X-ray edge problem.  Specifically, the first two terms represent the 
distinct scattering potential experienced by a ``metallic'' $a$-fermion whilst scattering off (Mott) localized $b$-fermion with same spin (first term in $H'$, $V=(U_{ab}+J/4)$) or with opposite spin (second term in $H'$, $V=(U_{ab}-J/4)$).  From the mapping onto the X-ray edge problem, it now follows that

$(i)$ the ``equal-spin excitonic'' fluctuation propagator, $\chi_{ab}^{\sigma\sigma}(\omega)=\int d\tau e^{i\omega\tau}\langle T_{tau}[a_{i\sigma}^{\dag}b_{i\sigma}(\tau);b_{i\sigma}^{\dag}a_{i\sigma}(0)]\rangle \simeq |\omega|^{-\eta_{1}}$, with $\eta_{1}=(2\delta_{1}/\pi -(\delta_{1}/\pi)^{2})$ and $\delta_{1}=$tan$^{-1}[(U_{ab}+J/4)\rho_{0}]$.

$(ii)$ the ``opposite-spin excitonic'' fluctuation propagator,  $\chi_{ab}^{\sigma,-\sigma}(\omega)=\int d\tau e^{i\omega\tau}\langle T_{tau}[a_{i\sigma}^{\dag}b_{i,-\sigma}(\tau);b_{i,-\sigma}^{\dag}a_{i\sigma}(0)]\rangle \simeq |\omega|^{-\eta_{2}}$, with $\eta_{2}=(2\delta_{2}/\pi -(\delta_{2}/\pi)^{2})$ and $\delta_{2}=$tan$^{-1}[(U_{ab}-J/4)\rho_{0}]$.

$(iii)$ the ``interband spin-flip excitonic'' fluctuation propagator, $\chi_{ab}^{sf}(\omega)= \int d\tau e^{i\omega\tau}\langle T_{tau}[a_{i\sigma}^{\dag}b_{i,\sigma}(\tau);b_{i,-\sigma}^{\dag}a_{i,-\sigma}(0)]\rangle \simeq |\omega|^{-\eta_{3}}$, with $\eta_{3}=(2\delta_{3}/\pi -(\delta_{3}/\pi)^{2})$ and $\delta_{3}=$tan$^{-1}(J\rho_{0}/2)$.

   The corresponding exponent in the spin fluctuation channel can be readily evaluated by using the above results and repeating the procedure detailed above for the charge fluctuation channel, but now with a different local scattering potential (related to $(ii),(iii)$ above).
   The different exponents in the power-law fall-off for the charge and spin susceptibilities found in our DMFT(CTQMC) study in the main text are thus rationalizable as arising from different scattering potentials experienced by the ``metallic'' $a$-fermions in the ``charge'' and ``spin'' fluctuation channels whilst scattering off the Mott localized $b$-states.  Obviously, the selective Mottness is a key factor in this emergent behavior, since it is only in this regime that the underlying impurity problem of DMFT maps onto the venerated X-ray edge problem, facilitating infra-red singular behavior.  Since DMFT is a self-consistently embedded single-impurity problem, the above singular behaviors carry over to the lattice problem, as long as one restricts oneself to the selective-metallic states without conventional symmetry breaking. 

   We thus arrive at one of our central results: the high-$D$ spin-charge separation alluded to in the main text arises from $(i)$ suppression of recoil of the ``heavy'' $b$-fermion during scattering processes (due to $U_{ab},J$) in the OSMP due to selective Mott localization, and $(ii)$ due to the different local scattering potentials (hence, different scattering phase shifts) in the charge and spin fluctuation sectors in the corresponding X-ray edge problem.

   Finally, for smaller $U_{ab}<U_{ab}^{OSMP}$, the one-electron hybrtidization, $V_{ab}(k)$ is relevant since, in the absence of selective Mott localization of the $b$-band fermions, the $b$-fermions can dynamically recoil
at low energies, leading to recovery of the lattice Kondo scale and to correlated Landau FL metallicity.  This is again in full qualitative accord with our DMFT(CTQMC) numerics.


\begin{thebibliography}{60}
\bibitem{civelli} M. Civelli, Ph.D Thesis (Rutgers Univ), arXiv:0710.2802.
\bibitem{aeppli}
G.~Aeppli and T.~Mason and S.~Hayden and H.~Mook and J.~Kulda,
\newblock Science {\bf 278}, 1432 (1997).
\bibitem{vdm} D van der Marel et al., 
\newblock Nature {\bf 425}, 271 (2003).
\bibitem{akrap} C. C. Homes et al., Nature Scientific Reports 3, Article number: 3446 (2013).
\bibitem{pwa} Philip W Anderson and Philip A Casey 2010 J. Phys.: Condens. Matter 22 164201.
\bibitem{qcp} M. Metlitski and S Sachdev, 
\newblock Phys Rev B {\bf 82}, 075127 (2010).
\bibitem{pepin} C. Pepin, Phys. Rev. B 77, 245129 (2008).
\bibitem{laad} M. S. Laad et al., Journal of Physics: Condensed Matter 24, 232204 (2012).
\bibitem{varma} C. M. Varma, Phys. Rev. B 73, 155113 (2006).
\bibitem{liviu} L. Hozoi et al., Phys Rev Lett. {\bf 99}, 256404 (2007).
\bibitem{imada} H. Sakakibara et al., Phys Rev B {\bf 89}, 224505 (2014).
\bibitem{swagata} S. Acharya {\it et al.}, Phys. Rev. X 8, 021038 (2018), M. S. Laad {\it et al.}, arXiv:0902.1903v1
\bibitem{giamarchi} C. Weber et al., Phys Rev Lett. {\bf 112}, 117001 (2014).
\bibitem{werner} Philipp Werner, Shintaro Hoshino, and Hiroshi Shinaoka,
Phys. Rev. B 94, 245134  (2016)
\bibitem{raghu} S. Raghu et al., 
\newblock Phys Rev B {\bf 77}, 220503(R), (2008).
\bibitem{hewson} A. Hewson, 
\newblock "The Kondo Problem to Heavy Fermions" Cambridge University Press (1993).
\bibitem{CT-HYB}
Philipp Werner, Armin Comanac, Luca de’ Medici, Matthias Troyer, and Andrew J. Millis,
\newblock Phys. Rev. Lett. 97, 076405 (2006).


\bibitem{ALPS}
\bibinfo{author}{Bauer, B.} \emph{et~al.}
\newblock \bibinfo{title}{The alps project release 2.0: open source software
  for strongly correlated systems}.
\newblock \emph{\bibinfo{journal}{Journal of Statistical Mechanics: Theory and
  Experiment}} \textbf{\bibinfo{volume}{2011}}, \bibinfo{pages}{P05001}
  (\bibinfo{year}{2011}).
  
  
\bibitem{sachdev-ye} N. Read et al., Phys Rev B {\bf 52}, 384 (1995).

\bibitem{mark} M. Jarrell, J. E. Gubernatis, and R. N. Silver, Phys. Rev. B
44, 5347 (Sep 1991). 
\bibitem{nsrep} S. Acharya et al., Scientific Reports volume 7, Article number: 43033 (2017), S. Acharya et al., J. Phys. Commun. https://doi.org/10.1088/2399-6528/aace29 (2018).
\bibitem{khveschenko} O. Parcollet and A. Georges, 
\newblock Phys Rev B {\bf 59}, 5341 (1999).
\bibitem{glossop} M. Glossop {\it et al.}, Phys Rev Lett. {\bf 107}, 076404 (2011).
\bibitem{schotte} K. D. Schotte and U. Schotte, Phys. Rev. 182, 479 (1969).
\bibitem{casey} Casey, P. A., Anderson, P. W. (2011). Physical review letters, 106(9), 097002.
%
%
%
%
%
\end{thebibliography}

\end{document}